\documentclass[twocolumn]{aastex631}
\usepackage{amsmath}     
\usepackage{wasysym}           
\usepackage{graphicx}
\usepackage{amssymb}
\usepackage{epstopdf}
\usepackage{mathrsfs}
\usepackage{anyfontsize}
\usepackage{natbib}
\usepackage{color}
\usepackage{lipsum}
\usepackage{diagbox}

\shorttitle{The TDE luminosity function}
\shortauthors{Coughlin \& Nicholl}
\begin{document}
\title{The luminosity function of TDEs from fallback-powered emission: implications for the black hole mass function}
\author[0000-0003-3765-6401]{Eric R.~Coughlin}
\affiliation{Department of Physics, Syracuse University, Syracuse, NY 13210, USA}
\author[0000-0002-2555-3192]{Matt Nicholl}
\affiliation{Astrophysics Research Centre, School of Mathematics and Physics, Queens University Belfast, Belfast BT7 1NN, UK}

\email{ecoughli@syr.edu}

\begin{abstract}
Tidal disruption events (TDEs), in which a star is destroyed by the gravitational field of a supermassive black hole (SMBH), are being observed at a high rate owing to the advanced state of survey science. One of the properties of TDEs that is measured with increasing statistical reliability is the TDE luminosity function, $d\dot{N}_{\rm TDE}/dL$, which is the TDE rate per luminosity (i.e., how many TDEs are within a given luminosity range). Here we show that if the luminous emission from a TDE is directly coupled to the rate of return of tidally destroyed debris to the SMBH, then the TDE luminosity function is in good agreement with observations and scales as $\propto L^{-2.5}$ for high luminosities, provided that the SMBH mass function $dN_{\bullet}/dM_{\bullet}$ -- the number of SMBHs ($N_{\bullet}$) per SMBH mass ($M_{\bullet}$) -- is approximately flat in the mass range over which we observe TDEs. We also show that there is a cutoff in the luminosity function at low luminosities that is a result of direct captures, and this cutoff has been tentatively observed. If $dN_{\bullet}/dM_{\bullet}$ is flat, which is in agreement with some observational campaigns, these results suggest that the fallback rate feeds the accretion rate in TDEs. Contrarily, if $dN_{\bullet}/d\log M_{\bullet}$ is flat, which has been found theoretically and is suggested by other observational investigations, then the emission from TDEs is likely powered by another mechanism. Future observations and more TDE statistics, provided by the Rubin Observatory/LSST, will provide additional evidence as to the reality of this tension.
\end{abstract}


\section{Introduction}
Time-domain surveys (e.g., ASAS-SN; \citealt{shappee14}; ZTF; \citealt{bellm19};  ATLAS; \citealt{tonry18}; PanSTARRS; \citealt{chambers16}; eROSITA; \citealt{predehl21}) have been discovering tidal disruption events (TDEs; e.g., \citealt{rees88, gezari21}) at an elevated rate, enabling an observational inference of TDE statistics, and this rate will only increase in the imminent era of the Rubin Observatory/LSST \citep{ivezic19,bricman20}. Among these TDE statistics is the TDE luminosity function, being the number of TDEs per unit time per unit luminosity {}{(where the luminosity is measured at the peak and in a given band, or an assumption is made to convert to a bolometric luminosity)}, or $d\dot{N}_{\rm TDE}/d L$. Recent samples of TDEs have been used to constrain this function: \cite{vanvelzen18} used a set of 12 TDEs and found that {}{the luminosity function of TDEs in the g-band was} $d\dot{N}_{\rm TDE}/dL \propto L^{-2.6 \pm 0.2}$. More recently, \citet{lin22} used a sample of 33 TDEs from the ZTF-I survey and found $d\dot{N}_{\rm TDE}/dL \propto L^{-2.3\pm0.2}$ {}{for g-band luminosities, or $\propto L^{-2.2\pm0.2}$ in terms of the bolometric luminosity} -- slightly shallower than the power-law slope inferred by \cite{vanvelzen18}. \citet{charal22} recovered {}{a g-band luminosity function of} $d\dot{N}_{\rm TDE}/dL \propto L^{-2.4}$ from a set of 30 TDEs from the ZTF catalog in \citet{hammerstein22}, and \citet{yao23} used 33 ZTF-obtained TDEs,{ finding that the g-band luminosity function favored $\propto L^{-3}$ when modeled as a single power-law, while the bolometric luminosity scaled as $\propto L^{-2.4}$}. \citet{yao23} also presented an observational SMBH mass function for their ZTF TDEs, obtained using the velocity dispersions in the host galaxies (i.e., derived independently of their luminosity function), finding an approximately flat distribution {}{(in $\log M_{\bullet}$ with $M_{\bullet}$ the supermassive black hole mass)} between $\sim10^{5}M_{\odot}$ and $10^{7}M_{\odot}${}{, which also agrees with the finding of \citet{vanvelzen18}}. 

On the other hand, most theoretical investigations have focused on the TDE rate per unit supermassive black hole (SMBH) mass, i.e., $d\dot{N}_{\rm TDE}/dM_{\bullet}$. \citet{magorrian99} and \citet{wang04} observationally constrained the loss cone filling rate -- the rate at which stars are fed into the region of angular momentum space that will take them within the tidal radius of the central SMBH (e.g., \citealt{frank76, lightman77, cohn78}) -- from observations of the central regions of nearby galaxies, which was then extended by \citet{stone16}. Given the loss cone filling rate, the rate at which (observable) TDEs occur is reduced owing to the possibility of direct capture \citep{hills75}: if the SMBH is especially massive ($\gtrsim 10^{7.5}M_{\odot}$ {}{for a solar-like star}), the tidal radius of the star can be within the direct capture radius of the black hole, equal to $4GM_{\bullet}/c^2$ for a non-spinning SMBH, thus rendering the TDE unobservable. Recent investigations by, e.g., \citet{beloborodov92, kesden12, will12, servin17, stone19} highlighted the fact that the direct capture radius can influence the TDE rate in a way that depends on black hole spin (on which the direct capture radius depends, as well as the projection of the angular momentum of the incoming star on the spin axis of the SMBH; see Equation 24 of \citealt{coughlin22b}), providing the possibility of constraining this parameter by using the precise variation of the TDE rate at the high-mass end of the SMBH mass function. However, there are additional factors that modify the direct capture condition that are related to, e.g., the population of tidally destroyed stars \citep{dorazio19}, and \citet{coughlin22b} pointed out that if the stellar mass function is dominated by low-mass stars, the cutoff in the TDE mass function should be closer to $10^{7}M_{\odot}$ rather than the often-quoted value of $\sim 10^{8}M_{\odot}$ (e.g., \citealt{hills75}). Modeling of observed TDEs suggests that essentially all are consistent with low mass stars and SMBH masses $\lesssim10^{7}M_{\odot}$ \citep{nicholl22}.

Connecting these two quantities -- the TDE rate per luminosity vs.~the TDE rate per SMBH mass -- requires an understanding of the light production mechanism(s) in TDEs and the relation between the black hole mass and the resultant TDE luminosity, which is an ongoing subject of debate (for various models see, e.g., \citealt{cannizzo90, loeb97, coughlin14, guillochon14, piran15, metzger16, roth16, bonnerot21, eyles22, metzger22}). However, the mass supply rate $\dot{M}$, or the rate at which tidally disrupted material returns to the SMBH -- usually referred to as the fallback rate -- depends on the SMBH mass as $\dot{M} \propto M_{\bullet}^{-1/2}$ (\citealt{lacy82, rees88}; see Figure 1 of \citealt{wu18} for numerical verification of this scaling; there are also dependencies on the stellar properties and the full vs.~partial nature of the disruption; e.g., \citealt{lodato09, guillochon13, mainetti17, golightly19a, lawsmith20, miles20, nixon21, coughlin22c}). How this fallback rate translates into the luminosity we ultimately observe is not yet clear, but perhaps the simplest prescription is to assume that the luminosity is proportional to the fallback rate (which must be correct to zeroth order, as obviously there will be no luminosity if the fallback rate is zero). This assumption is consistent with the results of \citet{mockler19} and \citet{nicholl22}, who found little to no viscous delay in the optically bright TDEs they modeled with MOSFiT \citep{guillochon18}. 

The question is therefore: if we adopt a one-to-one mapping between the TDE luminosity and the fallback rate, is the result consistent with the observed TDE luminosity function, and are there constraints we can place on the underlying SMBH mass distribution (given this fallback rate-luminosity prescription)? Here we perform this exercise in Section \ref{sec:luminosity} and show that, given an approximately flat SMBH mass distribution $dN_{\bullet}/dM_{\bullet}$ (at least over the range of SMBH masses that yield currently observed TDEs), we recover a luminosity function that is in excellent agreement with observations.
We also show that there is a cutoff in the TDE luminosity function at the low-luminosity end, which results from the direct capture radius of the SMBH (note that if the fallback rate powers the accretion luminosity, larger-mass SMBHs have smaller luminosities). We conclude and discuss the implications of our findings in Section \ref{sec:conclusions}.

\section{Theoretical TDE Luminosity function}
\label{sec:luminosity}
The conditional TDE rate (i.e., for a specified SMBH mass) is given by the product of the loss cone filling rate, $\dot{N}_{\rm lc}$, and the fraction of stars that enter the loss cone and are not directly captured, $N_{\rm tde}/N$, which is then integrated over the distribution function of tidally destroyed stars, $f_{\star}(M_{\star}, R_{\star})$. The loss cone filling rate has been analyzed extensively from the Fokker-Planck equation under the assumption that stars scatter in energy-angular momentum space through two-body interactions (e.g., \citealt{lightman77}), and upon using an $M$-$\sigma$ relation to eliminate the dependence on the velocity dispersion in the nucleus of a galaxy, \cite{merritt13} finds that the loss-cone filling rate is
\begin{equation}
\dot{N}_{\rm lc} \propto M_{\bullet}^{-0.25} \label{Lc}.
\end{equation}
This scaling (and relatively weak dependence on SMBH mass) is in agreement with the more recent results of \citet{stone16}. The fraction of tidally destroyed (and not directly captured) stars can then be estimated if the stellar distribution of specific angular momentum and -- if the SMBH has non-zero spin -- the projection of the angular momentum onto the SMBH spin axis are known. This fraction can be computed numerically \citep{kesden12}, but in general is an integral of the joint probability distribution function (of the square of the specific angular momentum and the projection onto the SMBH spin axis) over the direct capture region (in angular momentum space) of the SMBH. If stars are distributed isotropically at large distances from the SMBH and the loss cone is full, then for non-spinning SMBHs the fraction is particularly simple \citep{coughlin22b}:
\begin{equation}
\frac{N_{\rm TDE}}{N} = \left(1-\frac{4GM_{\bullet}/c^2}{r_{\rm t}}\right)^2 \label{ntde}.
\end{equation}
Here $r_{\rm t}$ is the tidal radius of the SMBH, which is often determined by equating the stellar surface gravity to the tidal field of the SMBH and dropping numerical factors, i.e.,
\begin{equation}
r_{\rm t} = R_{\star}\left(\frac{M_{\bullet}}{M_{\star}}\right)^{1/3}, \label{rtdef}
\end{equation}
where $R_{\star}$ and $M_{\star}$ are the stellar radius and mass. In general, however, an appropriate and more accurate definition can be used that incorporates more properties of the star. Equation \eqref{ntde} yields the expected results in the extreme limits of $r_{\rm t} = 4GM_{\bullet}/c^2$, where all stars are directly captured ($N_{\rm TDE}/N = 0$), and $r_{\rm t} \gg 4GM_{\bullet}/c^2$, where all stars are tidally disrupted and relativistic effects are ignorable ($N_{\rm TDE}/N \simeq 1$).

The TDE rate per unit SMBH mass per unit stellar mass (and radius) is then given by the product of the loss-cone filling rate, the fraction of tidally destroyed stars, the distribution function of stellar mass and radii, and the distribution function of SMBH mass, $dN_{\bullet}/dM_{\bullet}$. For concreteness, simplicity, and to emphasize the main result, we restrict ourselves to a given type of star (i.e., the distribution function of stellar mass and radii is a $\delta$-function, but see Section \ref{sec:conclusions} for additional discussion as to the consequences of relaxing this assumption), in which case the TDE rate per unit black hole mass is
\begin{equation}
\frac{d\dot{N}_{\rm tde}}{dM_{\bullet}} = \dot{N}_{\rm lc}\left(1-\frac{4GM_{\bullet}/c^2}{r_{\rm t}}\right)^2\frac{dN_{\bullet}}{dM_{\bullet}}. \label{Ndottde}
\end{equation}
As noted above, we need to convert between the SMBH mass and the corresponding TDE luminosity to derive the luminosity function. To do so, we make the simple assumption that the TDE luminosity scales in proportion to the usual TDE fallback rate:
\begin{equation}
L = \eta M_{\star}c^2\sqrt{GM_{\star}}R_{\star}^{-3/2}\left(\frac{M_{\bullet}}{M_{\star}}\right)^{-1/2}. \label{LofM}
\end{equation}
Here $\eta$ incorporates our ignorance of the emission mechanism(s) of TDEs, and could be close to the typical value of $0.1$ that is often invoked in accretion discs (e.g., \citealt{frank02}), or could be much smaller; observations tend to find that the total energy radiated in the optical/UV is at least 1-2 orders of magnitude smaller than the $\sim 10^{53}$ erg that would result from converting $0.5M_{\odot}$ of rest-mass energy into light at 10\% efficiency. This discrepancy suggests either that the majority of the energy is liberated in bands to which we are not sensitive, specifically the EUV \citep{lu18}, or the accretion process in a TDE is inherently inefficient due to the trapping and advection of radiation \citep{begelman78, coughlin14}, longer circularization timescales relative to the fallback time \citep{guillochon15}, and (or) the production of kinetic-energy-dominated outflows \citep{strubbe09, metzger16}. A lower-mass disrupted star would also reduce the overall energy budget. The precise value of $\eta$ will depend on the specific energy band under consideration, the type of star destroyed, the point of closest approach of the star to the SMBH (as partial disruptions yield both an overall smaller peak luminosity that is also shifted to later times; \citealt{guillochon13, miles20, nixon21}), and conceivably the SMBH mass. {}{For now we ignore these subtleties and let $\eta$ be a constant, but we discuss the implications of a $M_{\bullet}$-dependent $\eta$ in Section \ref{sec:conclusions}}. 

Under this approximation, Equation \eqref{LofM} can be immediately and algebraically rearranged to give $M_{\bullet} \propto L^{-2}$, and we have
\begin{equation}
\begin{split}
\frac{d\dot{N}_{\rm TDE}}{dL} &= \frac{dM_{\bullet}}{dL}\frac{d\dot{N}_{\rm TDE}}{dM_{\bullet}} \\ 
&\propto L^{-2.5}\left(1-\left(L/L_{\rm bk}\right)^{-4/3}\right)^2 \frac{dN_{\bullet}}{dM_{\bullet}}. \label{dNdL}
\end{split}
\end{equation}
Here the factor of $L^{-2.5}$ comes from the fact that $dM_{\bullet}/dL \propto L^{-3}$ and $\dot{N}_{\rm lc} \propto M_{\bullet}^{-0.25} \propto L^{0.5}$ (so specific assumptions about the loss-cone filling rate and the $M$-$\sigma$ relation could change this power-law by $\sim few\times 0.1$ dex), while the factor $(L/L_{\rm bk})^{-4/3}$ arises from combining the standard definition of the tidal radius (Equation \ref{rtdef}) with Equation \eqref{LofM}, i.e., 
\begin{equation}
\begin{split}
\frac{4GM_{\bullet}/c^2}{r_{\rm t}} &= \frac{4GM_{\star}}{c^2 R_{\star}}\left(\eta M_{\star}c^2\sqrt{GM_{\star}}R_{\star}^{-3/2}L^{-1}\right)^{4/3}  \\ 
&= \left(L/L_{\rm bk}\right)^{-4/3}, \\
L_{\rm bk} &\equiv \eta\left(\frac{4GM_{\star}}{c^2R_{\star}}\right)^{3/4}M_{\star}c^2\sqrt{GM_{\star}}R_{\star}^{-3/2}.
\end{split} \label{Lbk}
\end{equation}
With $\eta = 10^{-3}$ and solar-like values, we find $L_{\rm bk} \simeq 1.7\times 10^{44}$ erg s$^{-1}$, which is consistent with the location of an observed break in the power-law decline of the bolometric TDE luminosity function as found in \citet{yao23}. This small value of $\eta$ is also consistent with the low radiative efficiencies inferred through observations of optical- and UV-bright TDEs (see the discussion under Equation \ref{LofM}), but we emphasize that this value is not determined self-consistently here.

\section{Discussion and Conclusions}
\label{sec:conclusions}
\begin{figure}[htbp] 
   \centering
   \includegraphics[width=0.47\textwidth]{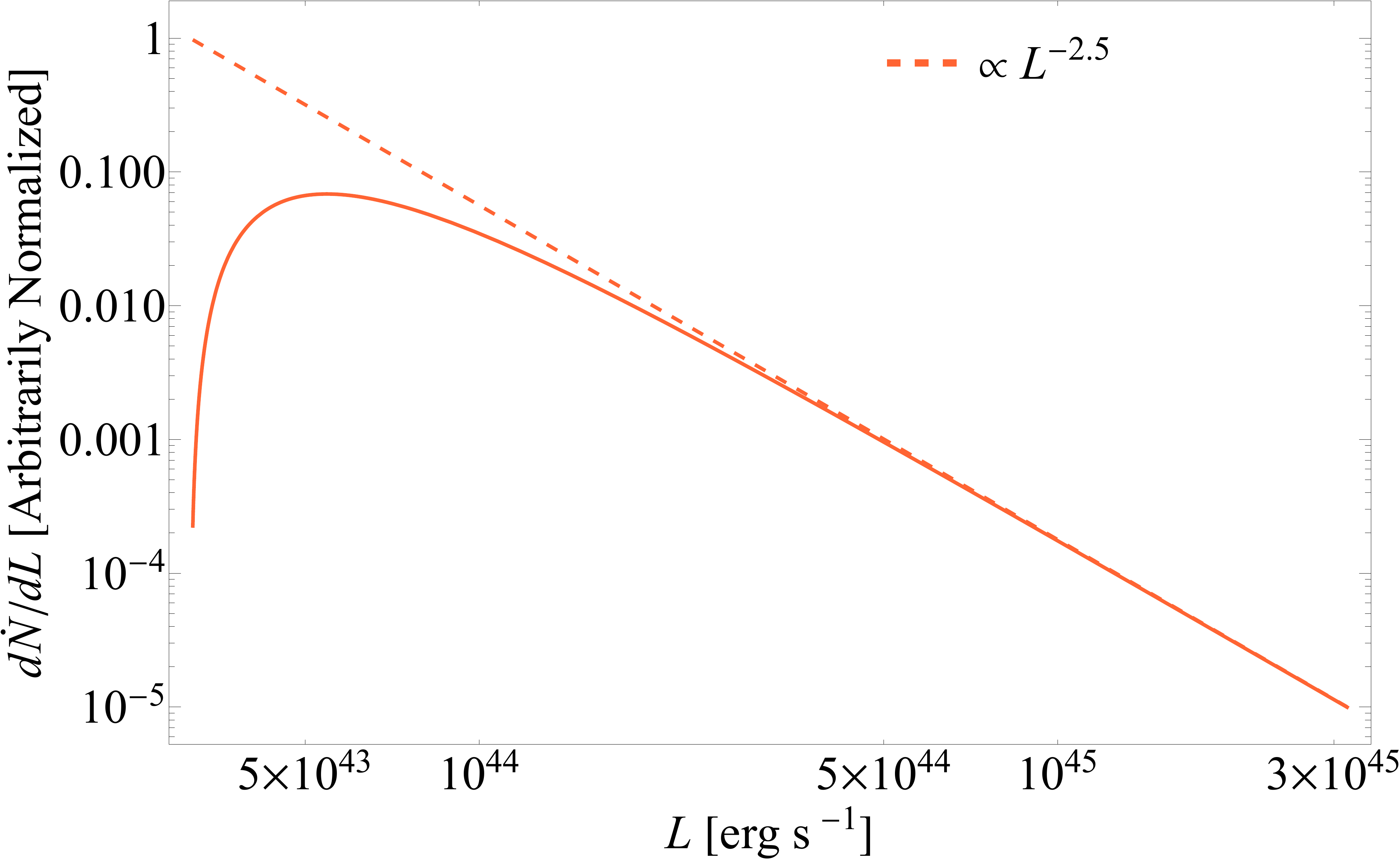} 
   \caption{The luminosity function as calculated in Equation \eqref{dNdL} with the break luminosity equal to $10^{43.5}$ erg s$^{-1}$. }
   \label{fig:lf}
\end{figure}

To highlight the functional form of Equation \eqref{dNdL}, Figure \ref{fig:lf} shows the luminosity function with $L_{\rm bk} = 10^{43.5}$ erg s$^{-1}$ and a constant SMBH mass function ($dN_{\bullet}/dM_{\bullet} = const.$). From this figure and the preceding calculation of the TDE luminosity function, which culminates in Equation \eqref{dNdL} (alongside Equation \ref{Lbk} for the limiting luminosity toward the lower end), we draw two conclusions:

\begin{enumerate}
\item If $dN_{\bullet}/dM_{\bullet} \simeq const.${}{, i.e., if the SMBH mass function --- the number of SMBHs per unit SMBH mass --- is uniform}, then the TDE luminosity function falls off as $d\dot{N}_{\rm TDE}/dL \propto L^{-2.5}$ {}{for $L \gtrsim L_{\rm bk}$}. This scaling is in remarkably good agreement with observational results, namely \citet{vanvelzen18, charal22, lin22, yao23}. 
\item There is a luminosity below which we do not expect to find TDEs, which corresponds to $L_{\rm bk}$ as given in Equation \eqref{Lbk}. This cutoff in the luminosity function arises from the direct capture radius of a SMBH and the inverse relationship between the fallback luminosity and the black hole mass -- since larger SMBH masses correspond to smaller luminosities according to the fallback rate, the direct capture of stars suppresses TDEs at the low-luminosity end of the distribution. This suppression was found observationally in \citet{lin22}, but those authors attributed this cutoff to the Eddington luminosity of the black hole. 
\end{enumerate}

{Agreement between Equation \eqref{dNdL} and the observed luminosity function of TDEs is satisfied if  $dN_{\bullet}/dM_{\bullet} \simeq const.$ over the range of SMBH masses at which we observe TDEs (generally expected to be toward the low-mass end), and the constancy of the SMBH mass function at low masses ($\lesssim 10^{7.5}M_{\odot}$) is roughly consistent with the observational results of \cite{graham07}. In particular, by parameterizing $dN_{\bullet}/dM_{\bullet} \propto M_{\bullet}^{\alpha}$, they find $\alpha \simeq 0\pm0.1$ for early-type galaxies, while $\alpha \simeq -0.3$ over their entire galaxy sample (see their Table 3). A $\sim$ flat distribution of $dN_{\bullet}/dM_{\bullet}$ at low masses was also found by \cite{davis14} for late-type galaxies (see their Figures 8 and 10), and similarly for the work of \cite{mutlu16} (see their Figure 10).} 

{On the other hand, earlier observational work (e.g., \citealt{shankar04, marconi04}) and theoretical investigations seem to favor $dN_{\bullet}/d\log M_{\bullet} \simeq const.$ (or weakly diverging with decreasing mass), implying $dN_{\bullet}/dM_{\bullet} \propto M_{\bullet}^{-1}$ and thus a strongly diverging mass function at small SMBH masses (e.g., \citealt{hopkins07, merloni08, shankar09, volonteri10}; see the review by \citealt{kelly12}; see also Figure 1 of \citealt{kochanek16}). With a SMBH mass function that is flat in $\log M_{\bullet}$, we have $dN_{\bullet}/dM_{\bullet} \propto 1/M_{\bullet} \propto L^{2}$, implying that $d\dot{N}_{\rm TDE}/dL \propto L^{-0.5}$. Such a weak scaling with luminosity is not consistent with observations.}

{There are a few possibilities that would reconcile this discrepancy, one of which is that the theoretical estimates of the SMBH mass function over-represent the number of SMBHs at small mass, or empirical SMBH relations (e.g., $M-\sigma$) are not valid toward the low-mass end of the mass function. Another is that the fallback rate simply does not scale with the accretion luminosity of the SMBH. This would not be surprising on a case-by-case basis owing to the (likely) complexity of circularization, viscous delays, and the potential production of outflows, which would produce a distinct observational signature (e.g., \citealt{strubbe09}; see also the various models considered in Appendix B of \citealt{stone16}).  Nevertheless, the fallback rate seems to track the accretion luminosity in optically bright TDEs with little to no viscous delay \citep{mockler19, nicholl22}.}

{It is also possible that there is a strong correlation between the type of star disrupted and the SMBH mass. As pointed out by \citet{kochanek16}, the dominance of low-mass stars by number in the stellar mass function implies that they should contribute predominantly to the TDE rate for small SMBH mass. This is also consistent with the results of \citet{mockler19, nicholl22}, who find that low-mass stars tend to fit optical TDE lightcurves when modeled with MOSFiT. However, for larger-mass SMBHs, low-mass stars are directly captured, meaning that only high-mass stars can contribute to the observable TDE rate; this rate suppression is directly incorporated into the luminosity function via the term $\left(1-4GM_{\bullet}/c^2/r_{\rm t}\right)^2$. There is also a non-trivial dependence on the scaling of the luminosity with the properties of the star, and both numerical simulations \citep{lawsmith20, ryu20} and physical arguments \citep{coughlin22c} suggest that the bulk stellar parameters in Equation \eqref{LofM} do not capture the entirety of the behavior. However, if we adopt a mass-radius relationship of the form $R_{\star} \propto M_{\star}$, approximately valid for low-mass stars \citep{demircan91}, then Equation \eqref{LofM} shows $L \propto \left(M_{\bullet}/M_{\star}\right)^{-1/2}$, while $M_{\bullet}/r_{\rm t} \propto \left(M_{\bullet}/M_{\star}\right)^{2/3}$. In this case, then, the integration over the stellar mass function does not introduce any additional dependence on the luminosity.}

{It may also be that the radiative efficiency -- $\eta$ in Equation \eqref{LofM} -- is a function of the SMBH mass. For example, high-mass SMBHs have more relativistic tidal radii for a given stellar type, which would facilitate circularization and the dissipation of energy. Since the relativistic apsidal precession angle scales with $M_{\bullet}$ to lowest order in $GM_{\bullet}/(r_{\rm t}c^2)$, we might naively expect $\eta \propto M_{\bullet}$. The Eddington limit may also cap the luminosity for low-mass SMBHs (although jetted TDEs may overcome this barrier at the expense of jet production; e.g., \citealt{zauderer13, pasham23}), for which $L \propto M_{\bullet}$ and hence $\eta \propto M_{\bullet}^{3/2}$. In either case the scaling with the luminosity would correlate positively (rather than inversely) with the SMBH mass, which would result in a much shallower TDE luminosity function or even an increasing one, neither of which is observed.}

There are other assumptions that we made to simplify the analysis and arrive at Equation \eqref{dNdL}, e.g., we neglected partial disruptions and the additional dependence on the impact parameter, we ignored the possible non-zero SMBH spin, and we set the loss-cone filling rate to $\dot{N}_{\rm lc} \propto M_{\bullet}^{-0.25}$. The relaxation of these assumptions would likely result in, e.g., a smearing out of the break luminosity $L_{\rm bk}$ in the TDE luminosity function, as the direct capture radius of the SMBH depends on both the SMBH {mass and} spin, and this reintroduces a non-trivial dependence of the direct capture radius on the stellar type. Variation in the loss cone filling rate would also modify and induce scatter in the power-law decline of the luminosity function. Some of these assumptions can be fairly straightforwardly relaxed by adopting a given stellar population, parameterizing the fallback rate as a function of the point of closest approach of the star to the black hole (primarily to account for partial disruptions, the fallback rates from which display a strong dependence on this parameter; e.g., \citealt{guillochon13, miles20, lawsmith20, nixon21}), and including the likelihood of reaching a given pericenter distance. For example, if most luminous TDEs originate from the full loss cone where the distribution of the square of the specific stellar angular momentum is uniform, then the probability distribution function of the pericenter distance is (Equation 10 of \citealt{coughlin22b})
\begin{equation}
f_{\rm r_{\rm p}}(r_{\rm p})\propto \frac{r_{\rm p}\left(r_{\rm p}-4\right)}{\left(r_{\rm p}-2\right)^2} \label{frp}
\end{equation}
for a Schwarzschild black hole. 

Including these refinements in the predicted TDE luminosity function alongside a large TDE population from LSST, the luminosity function around and below $L_{\rm bk}$ may provide a means to statistically constrain SMBH spins, because -- as we noted in the preceding paragraph -- the direct capture radius and the luminosity below which TDEs are no longer expected to occur (which impacts the luminosity function through the fraction of stars that enter the loss cone and are disrupted, $N_{\rm TDE}/N$; see Equation \ref{ntde} above for the Schwarzschild value) depends on the SMBH spin. The pericenter distribution function, which is given by Equation \ref{frp} for a non-spinning SMBH, also depends on the SMBH spin in a non-trivial way that is, nonetheless, able to be determined straightforwardly in the full-loss-cone limit (see Figure 3 of \citealt{coughlin22b}). Parameterizing the distribution of SMBH spins, and incorporating the spin dependence into the TDE luminosity function through these estimates, would then serve to elucidate the spin-dependent behavior and shape of the luminosity function around the peak. Comparisons between the predicted luminosity function near the peak and a larger census of TDEs obtained by LSST could then serve to constrain SMBH spins in a statistically meaningful way, provided that the stellar population responsible for generating TDEs can be accurately determined. 

Nevertheless, in its simplest possible form, this model -- which proposes a one-to-one correspondence between fallback accretion and TDE emission -- reproduces the TDE luminosity function extremely well if the SMBH mass function $dN_{\bullet}/dM_{\bullet}$ is flat, and the latter agrees with some observational and theoretical work. On the other hand, if the SMBH mass function is flat in $\log M_{\bullet}$, which is suggested by a number of theoretical and other observational results, then there is a large discrepancy between this prediction and the observed luminosity function. In the latter case, the property of TDEs that is arguably best understood -- the fallback rate -- is sub-dominant to other, relatively unknown mechanisms for producing the luminous emission from TDEs. The accumulation of more TDE statistics with LSST, ZTF, and ASAS-SN, alongside independent constraints of the SMBH mass function toward the low-mass end, will shed light on which of these two scenarios is favored.
\newline
\newline
\indent{}We thank Chris Nixon and D.J.~Pasham for useful discussions. We also thank the anonymous referee for a useful and constructive report. E.R.C.~acknowledges support from the National Science Foundation through grant AST-2006684, and a Ralph E. Powe Junior Faculty Enhancement Award through the Oakridge Associated Universities. M.N.~is supported by the European Research Council (ERC) under the European Union’s Horizon 2020 research and innovation programme (grant agreement No.~948381) and by funding from the UK Space Agency.  Part of the inspiration for this work arose from discussions following the eXtreme Black Holes conference at the Aspen Center for Physics (which is supported by National Science Foundation grant PHY-2210452), particularly during a (long) car ride from Aspen to Denver after many canceled flights.

\bibliographystyle{aasjournal}


\begin{thebibliography}{}
\expandafter\ifx\csname natexlab\endcsname\relax\def\natexlab#1{#1}\fi
\providecommand{\url}[1]{\href{#1}{#1}}
\providecommand{\dodoi}[1]{doi:~\href{http://doi.org/#1}{\nolinkurl{#1}}}
\providecommand{\doeprint}[1]{\href{http://ascl.net/#1}{\nolinkurl{http://ascl.net/#1}}}
\providecommand{\doarXiv}[1]{\href{https://arxiv.org/abs/#1}{\nolinkurl{https://arxiv.org/abs/#1}}}

\bibitem[{{Begelman}(1978)}]{begelman78}
{Begelman}, M.~C. 1978, \mnras, 184, 53, \dodoi{10.1093/mnras/184.1.53}

\bibitem[{{Bellm} {et~al.}(2019){Bellm}, {Kulkarni}, {Graham}, {Dekany},
  {Smith}, {Riddle}, {Masci}, {Helou}, {Prince}, {Adams}, {Barbarino},
  {Barlow}, {Bauer}, {Beck}, {Belicki}, {Biswas}, {Blagorodnova}, {Bodewits},
  {Bolin}, {Brinnel}, {Brooke}, {Bue}, {Bulla}, {Burruss}, {Cenko}, {Chang},
  {Connolly}, {Coughlin}, {Cromer}, {Cunningham}, {De}, {Delacroix}, {Desai},
  {Duev}, {Eadie}, {Farnham}, {Feeney}, {Feindt}, {Flynn}, {Franckowiak},
  {Frederick}, {Fremling}, {Gal-Yam}, {Gezari}, {Giomi}, {Goldstein},
  {Golkhou}, {Goobar}, {Groom}, {Hacopians}, {Hale}, {Henning}, {Ho}, {Hover},
  {Howell}, {Hung}, {Huppenkothen}, {Imel}, {Ip}, {Ivezi{\'c}}, {Jackson},
  {Jones}, {Juric}, {Kasliwal}, {Kaspi}, {Kaye}, {Kelley}, {Kowalski},
  {Kramer}, {Kupfer}, {Landry}, {Laher}, {Lee}, {Lin}, {Lin}, {Lunnan},
  {Giomi}, {Mahabal}, {Mao}, {Miller}, {Monkewitz}, {Murphy}, {Ngeow},
  {Nordin}, {Nugent}, {Ofek}, {Patterson}, {Penprase}, {Porter}, {Rauch},
  {Rebbapragada}, {Reiley}, {Rigault}, {Rodriguez}, {van Roestel}, {Rusholme},
  {van Santen}, {Schulze}, {Shupe}, {Singer}, {Soumagnac}, {Stein}, {Surace},
  {Sollerman}, {Szkody}, {Taddia}, {Terek}, {Van Sistine}, {van Velzen},
  {Vestrand}, {Walters}, {Ward}, {Ye}, {Yu}, {Yan}, \& {Zolkower}}]{bellm19}
{Bellm}, E.~C., {Kulkarni}, S.~R., {Graham}, M.~J., {et~al.} 2019, \pasp, 131,
  018002, \dodoi{10.1088/1538-3873/aaecbe}

\bibitem[{{Beloborodov} {et~al.}(1992){Beloborodov}, {Illarionov}, {Ivanov}, \&
  {Polnarev}}]{beloborodov92}
{Beloborodov}, A.~M., {Illarionov}, A.~F., {Ivanov}, P.~B., \& {Polnarev},
  A.~G. 1992, \mnras, 259, 209, \dodoi{10.1093/mnras/259.2.209}

\bibitem[{{Bonnerot} {et~al.}(2021){Bonnerot}, {Lu}, \& {Hopkins}}]{bonnerot21}
{Bonnerot}, C., {Lu}, W., \& {Hopkins}, P.~F. 2021, \mnras, 504, 4885,
  \dodoi{10.1093/mnras/stab398}

\bibitem[{{Bricman} \& {Gomboc}(2020)}]{bricman20}
{Bricman}, K., \& {Gomboc}, A. 2020, \apj, 890, 73,
  \dodoi{10.3847/1538-4357/ab6989}

\bibitem[{{Cannizzo} {et~al.}(1990){Cannizzo}, {Lee}, \&
  {Goodman}}]{cannizzo90}
{Cannizzo}, J.~K., {Lee}, H.~M., \& {Goodman}, J. 1990, \apj, 351, 38,
  \dodoi{10.1086/168442}

\bibitem[{{Chambers} {et~al.}(2016){Chambers}, {Magnier}, {Metcalfe},
  {Flewelling}, {Huber}, {Waters}, {Denneau}, {Draper}, {Farrow}, {Finkbeiner},
  {Holmberg}, {Koppenhoefer}, {Price}, {Rest}, {Saglia}, {Schlafly}, {Smartt},
  {Sweeney}, {Wainscoat}, {Burgett}, {Chastel}, {Grav}, {Heasley}, {Hodapp},
  {Jedicke}, {Kaiser}, {Kudritzki}, {Luppino}, {Lupton}, {Monet}, {Morgan},
  {Onaka}, {Shiao}, {Stubbs}, {Tonry}, {White}, {Ba{\~n}ados}, {Bell},
  {Bender}, {Bernard}, {Boegner}, {Boffi}, {Botticella}, {Calamida},
  {Casertano}, {Chen}, {Chen}, {Cole}, {Deacon}, {Frenk}, {Fitzsimmons},
  {Gezari}, {Gibbs}, {Goessl}, {Goggia}, {Gourgue}, {Goldman}, {Grant},
  {Grebel}, {Hambly}, {Hasinger}, {Heavens}, {Heckman}, {Henderson}, {Henning},
  {Holman}, {Hopp}, {Ip}, {Isani}, {Jackson}, {Keyes}, {Koekemoer}, {Kotak},
  {Le}, {Liska}, {Long}, {Lucey}, {Liu}, {Martin}, {Masci}, {McLean}, {Mindel},
  {Misra}, {Morganson}, {Murphy}, {Obaika}, {Narayan}, {Nieto-Santisteban},
  {Norberg}, {Peacock}, {Pier}, {Postman}, {Primak}, {Rae}, {Rai}, {Riess},
  {Riffeser}, {Rix}, {R{\"o}ser}, {Russel}, {Rutz}, {Schilbach}, {Schultz},
  {Scolnic}, {Strolger}, {Szalay}, {Seitz}, {Small}, {Smith}, {Soderblom},
  {Taylor}, {Thomson}, {Taylor}, {Thakar}, {Thiel}, {Thilker}, {Unger},
  {Urata}, {Valenti}, {Wagner}, {Walder}, {Walter}, {Watters}, {Werner},
  {Wood-Vasey}, \& {Wyse}}]{chambers16}
{Chambers}, K.~C., {Magnier}, E.~A., {Metcalfe}, N., {et~al.} 2016, arXiv
  e-prints, arXiv:1612.05560, \dodoi{10.48550/arXiv.1612.05560}

\bibitem[{{Charalampopoulos} {et~al.}(2022){Charalampopoulos}, {Pursiainen},
  {Leloudas}, {Arcavi}, {Newsome}, {Schulze}, {Burke}, \& {Nicholl}}]{charal22}
{Charalampopoulos}, P., {Pursiainen}, M., {Leloudas}, G., {et~al.} 2022, arXiv
  e-prints, arXiv:2209.12913, \dodoi{10.48550/arXiv.2209.12913}

\bibitem[{{Cohn} \& {Kulsrud}(1978)}]{cohn78}
{Cohn}, H., \& {Kulsrud}, R.~M. 1978, \apj, 226, 1087, \dodoi{10.1086/156685}

\bibitem[{{Coughlin} \& {Begelman}(2014)}]{coughlin14}
{Coughlin}, E.~R., \& {Begelman}, M.~C. 2014, \apj, 781, 82,
  \dodoi{10.1088/0004-637X/781/2/82}

\bibitem[{{Coughlin} \& {Nixon}(2022{\natexlab{a}})}]{coughlin22b}
{Coughlin}, E.~R., \& {Nixon}, C.~J. 2022{\natexlab{a}}, \apj, 936, 70,
  \dodoi{10.3847/1538-4357/ac85b3}

\bibitem[{{Coughlin} \& {Nixon}(2022{\natexlab{b}})}]{coughlin22c}
---. 2022{\natexlab{b}}, \mnras, 517, L26, \dodoi{10.1093/mnrasl/slac106}

\bibitem[{{Davis} {et~al.}(2014){Davis}, {Berrier}, {Johns}, {Shields},
  {Hartley}, {Kennefick}, {Kennefick}, {Seigar}, \& {Lacy}}]{davis14}
{Davis}, B.~L., {Berrier}, J.~C., {Johns}, L., {et~al.} 2014, \apj, 789, 124,
  \dodoi{10.1088/0004-637X/789/2/124}

\bibitem[{{Demircan} \& {Kahraman}(1991)}]{demircan91}
{Demircan}, O., \& {Kahraman}, G. 1991, \apss, 181, 313,
  \dodoi{10.1007/BF00639097}

\bibitem[{{D'Orazio} {et~al.}(2019){D'Orazio}, {Loeb}, \&
  {Guillochon}}]{dorazio19}
{D'Orazio}, D.~J., {Loeb}, A., \& {Guillochon}, J. 2019, \mnras, 485, 4413,
  \dodoi{10.1093/mnras/stz652}

\bibitem[{{Eyles-Ferris} {et~al.}(2022){Eyles-Ferris}, {Starling}, {O'Brien},
  {Nixon}, \& {Coughlin}}]{eyles22}
{Eyles-Ferris}, R.~A.~J., {Starling}, R.~L.~C., {O'Brien}, P.~T., {Nixon},
  C.~J., \& {Coughlin}, E.~R. 2022, \mnras, 517, 6013,
  \dodoi{10.1093/mnras/stac3073}

\bibitem[{{Frank} {et~al.}(2002){Frank}, {King}, \& {Raine}}]{frank02}
{Frank}, J., {King}, A., \& {Raine}, D.~J. 2002, {Accretion Power in
  Astrophysics: Third Edition}

\bibitem[{{Frank} \& {Rees}(1976)}]{frank76}
{Frank}, J., \& {Rees}, M.~J. 1976, \mnras, 176, 633,
  \dodoi{10.1093/mnras/176.3.633}

\bibitem[{{Gezari}(2021)}]{gezari21}
{Gezari}, S. 2021, \araa, 59, 21, \dodoi{10.1146/annurev-astro-111720-030029}

\bibitem[{{Golightly} {et~al.}(2019){Golightly}, {Nixon}, \&
  {Coughlin}}]{golightly19a}
{Golightly}, E.~C.~A., {Nixon}, C.~J., \& {Coughlin}, E.~R. 2019, \apjl, 882,
  L26, \dodoi{10.3847/2041-8213/ab380d}

\bibitem[{{Graham} {et~al.}(2007){Graham}, {Driver}, {Allen}, \&
  {Liske}}]{graham07}
{Graham}, A.~W., {Driver}, S.~P., {Allen}, P.~D., \& {Liske}, J. 2007, \mnras,
  378, 198, \dodoi{10.1111/j.1365-2966.2007.11770.x}

\bibitem[{{Guillochon} {et~al.}(2014){Guillochon}, {Manukian}, \&
  {Ramirez-Ruiz}}]{guillochon14}
{Guillochon}, J., {Manukian}, H., \& {Ramirez-Ruiz}, E. 2014, \apj, 783, 23,
  \dodoi{10.1088/0004-637X/783/1/23}

\bibitem[{{Guillochon} {et~al.}(2018){Guillochon}, {Nicholl}, {Villar},
  {Mockler}, {Narayan}, {Mandel}, {Berger}, \& {Williams}}]{guillochon18}
{Guillochon}, J., {Nicholl}, M., {Villar}, V.~A., {et~al.} 2018, \apjs, 236, 6,
  \dodoi{10.3847/1538-4365/aab761}

\bibitem[{{Guillochon} \& {Ramirez-Ruiz}(2013)}]{guillochon13}
{Guillochon}, J., \& {Ramirez-Ruiz}, E. 2013, \apj, 767, 25,
  \dodoi{10.1088/0004-637X/767/1/25}

\bibitem[{{Guillochon} \& {Ramirez-Ruiz}(2015)}]{guillochon15}
---. 2015, \apj, 809, 166, \dodoi{10.1088/0004-637X/809/2/166}

\bibitem[{{Hammerstein} {et~al.}(2022){Hammerstein}, {van Velzen}, {Gezari},
  {Cenko}, {Yao}, {Ward}, {Frederick}, {Villanueva}, {Somalwar}, {Graham},
  {Kulkarni}, {Stern}, {Bellm}, {Dekany}, {Drake}, {Groom}, {Kasliwal}, {Kool},
  {Masci}, {Medford}, \& {van Roestel}}]{hammerstein22}
{Hammerstein}, E., {van Velzen}, S., {Gezari}, S., {et~al.} 2022, arXiv
  e-prints, arXiv:2203.01461.
\newblock \doarXiv{2203.01461}

\bibitem[{{Hills}(1975)}]{hills75}
{Hills}, J.~G. 1975, \nat, 254, 295, \dodoi{10.1038/254295a0}

\bibitem[{{Hopkins} {et~al.}(2007){Hopkins}, {Richards}, \&
  {Hernquist}}]{hopkins07}
{Hopkins}, P.~F., {Richards}, G.~T., \& {Hernquist}, L. 2007, \apj, 654, 731,
  \dodoi{10.1086/509629}

\bibitem[{{Ivezi{\'c}} {et~al.}(2019){Ivezi{\'c}}, {Kahn}, {Tyson}, {Abel},
  {Acosta}, {Allsman}, {Alonso}, {AlSayyad}, {Anderson}, {Andrew}, {Angel},
  {Angeli}, {Ansari}, {Antilogus}, {Araujo}, {Armstrong}, {Arndt}, {Astier},
  {Aubourg}, {Auza}, {Axelrod}, {Bard}, {Barr}, {Barrau}, {Bartlett}, {Bauer},
  {Bauman}, {Baumont}, {Bechtol}, {Bechtol}, {Becker}, {Becla}, {Beldica},
  {Bellavia}, {Bianco}, {Biswas}, {Blanc}, {Blazek}, {Blandford}, {Bloom},
  {Bogart}, {Bond}, {Booth}, {Borgland}, {Borne}, {Bosch}, {Boutigny},
  {Brackett}, {Bradshaw}, {Brandt}, {Brown}, {Bullock}, {Burchat}, {Burke},
  {Cagnoli}, {Calabrese}, {Callahan}, {Callen}, {Carlin}, {Carlson},
  {Chandrasekharan}, {Charles-Emerson}, {Chesley}, {Cheu}, {Chiang}, {Chiang},
  {Chirino}, {Chow}, {Ciardi}, {Claver}, {Cohen-Tanugi}, {Cockrum}, {Coles},
  {Connolly}, {Cook}, {Cooray}, {Covey}, {Cribbs}, {Cui}, {Cutri}, {Daly},
  {Daniel}, {Daruich}, {Daubard}, {Daues}, {Dawson}, {Delgado}, {Dellapenna},
  {de Peyster}, {de Val-Borro}, {Digel}, {Doherty}, {Dubois},
  {Dubois-Felsmann}, {Durech}, {Economou}, {Eifler}, {Eracleous}, {Emmons},
  {Fausti Neto}, {Ferguson}, {Figueroa}, {Fisher-Levine}, {Focke}, {Foss},
  {Frank}, {Freemon}, {Gangler}, {Gawiser}, {Geary}, {Gee}, {Geha}, {Gessner},
  {Gibson}, {Gilmore}, {Glanzman}, {Glick}, {Goldina}, {Goldstein}, {Goodenow},
  {Graham}, {Gressler}, {Gris}, {Guy}, {Guyonnet}, {Haller}, {Harris},
  {Hascall}, {Haupt}, {Hernandez}, {Herrmann}, {Hileman}, {Hoblitt}, {Hodgson},
  {Hogan}, {Howard}, {Huang}, {Huffer}, {Ingraham}, {Innes}, {Jacoby}, {Jain},
  {Jammes}, {Jee}, {Jenness}, {Jernigan}, {Jevremovi{\'c}}, {Johns}, {Johnson},
  {Johnson}, {Jones}, {Juramy-Gilles}, {Juri{\'c}}, {Kalirai}, {Kallivayalil},
  {Kalmbach}, {Kantor}, {Karst}, {Kasliwal}, {Kelly}, {Kessler}, {Kinnison},
  {Kirkby}, {Knox}, {Kotov}, {Krabbendam}, {Krughoff}, {Kub{\'a}nek},
  {Kuczewski}, {Kulkarni}, {Ku}, {Kurita}, {Lage}, {Lambert}, {Lange},
  {Langton}, {Le Guillou}, {Levine}, {Liang}, {Lim}, {Lintott}, {Long},
  {Lopez}, {Lotz}, {Lupton}, {Lust}, {MacArthur}, {Mahabal}, {Mandelbaum},
  {Markiewicz}, {Marsh}, {Marshall}, {Marshall}, {May}, {McKercher}, {McQueen},
  {Meyers}, {Migliore}, {Miller}, {Mills}, {Miraval}, {Moeyens}, {Moolekamp},
  {Monet}, {Moniez}, {Monkewitz}, {Montgomery}, {Morrison}, {Mueller},
  {Muller}, {Mu{\~n}oz Arancibia}, {Neill}, {Newbry}, {Nief}, {Nomerotski},
  {Nordby}, {O'Connor}, {Oliver}, {Olivier}, {Olsen}, {O'Mullane}, {Ortiz},
  {Osier}, {Owen}, {Pain}, {Palecek}, {Parejko}, {Parsons}, {Pease},
  {Peterson}, {Peterson}, {Petravick}, {Libby Petrick}, {Petry},
  {Pierfederici}, {Pietrowicz}, {Pike}, {Pinto}, {Plante}, {Plate}, {Plutchak},
  {Price}, {Prouza}, {Radeka}, {Rajagopal}, {Rasmussen}, {Regnault}, {Reil},
  {Reiss}, {Reuter}, {Ridgway}, {Riot}, {Ritz}, {Robinson}, {Roby}, {Roodman},
  {Rosing}, {Roucelle}, {Rumore}, {Russo}, {Saha}, {Sassolas}, {Schalk},
  {Schellart}, {Schindler}, {Schmidt}, {Schneider}, {Schneider}, {Schoening},
  {Schumacher}, {Schwamb}, {Sebag}, {Selvy}, {Sembroski}, {Seppala}, {Serio},
  {Serrano}, {Shaw}, {Shipsey}, {Sick}, {Silvestri}, {Slater}, {Smith},
  {Smith}, {Sobhani}, {Soldahl}, {Storrie-Lombardi}, {Stover}, {Strauss},
  {Street}, {Stubbs}, {Sullivan}, {Sweeney}, {Swinbank}, {Szalay}, {Takacs},
  {Tether}, {Thaler}, {Thayer}, {Thomas}, {Thornton}, {Thukral}, {Tice},
  {Trilling}, {Turri}, {Van Berg}, {Vanden Berk}, {Vetter}, {Virieux},
  {Vucina}, {Wahl}, {Walkowicz}, {Walsh}, {Walter}, {Wang}, {Wang}, {Warner},
  {Wiecha}, {Willman}, {Winters}, {Wittman}, {Wolff}, {Wood-Vasey}, {Wu},
  {Xin}, {Yoachim}, \& {Zhan}}]{ivezic19}
{Ivezi{\'c}}, {\v{Z}}., {Kahn}, S.~M., {Tyson}, J.~A., {et~al.} 2019, \apj,
  873, 111, \dodoi{10.3847/1538-4357/ab042c}

\bibitem[{{Kelly} \& {Merloni}(2012)}]{kelly12}
{Kelly}, B.~C., \& {Merloni}, A. 2012, Advances in Astronomy, 2012, 970858,
  \dodoi{10.1155/2012/970858}

\bibitem[{{Kesden}(2012)}]{kesden12}
{Kesden}, M. 2012, \prd, 85, 024037, \dodoi{10.1103/PhysRevD.85.024037}

\bibitem[{{Kochanek}(2016)}]{kochanek16}
{Kochanek}, C.~S. 2016, \mnras, 461, 371, \dodoi{10.1093/mnras/stw1290}

\bibitem[{{Lacy} {et~al.}(1982){Lacy}, {Townes}, \& {Hollenbach}}]{lacy82}
{Lacy}, J.~H., {Townes}, C.~H., \& {Hollenbach}, D.~J. 1982, \apj, 262, 120,
  \dodoi{10.1086/160402}

\bibitem[{{Law-Smith} {et~al.}(2020){Law-Smith}, {Coulter}, {Guillochon},
  {Mockler}, \& {Ramirez-Ruiz}}]{lawsmith20}
{Law-Smith}, J. A.~P., {Coulter}, D.~A., {Guillochon}, J., {Mockler}, B., \&
  {Ramirez-Ruiz}, E. 2020, \apj, 905, 141, \dodoi{10.3847/1538-4357/abc489}

\bibitem[{{Lightman} \& {Shapiro}(1977)}]{lightman77}
{Lightman}, A.~P., \& {Shapiro}, S.~L. 1977, \apj, 211, 244,
  \dodoi{10.1086/154925}

\bibitem[{{Lin} {et~al.}(2022){Lin}, {Jiang}, {Kong}, {Huang}, {Lin}, {Zhu}, \&
  {Wang}}]{lin22}
{Lin}, Z., {Jiang}, N., {Kong}, X., {et~al.} 2022, \apjl, 939, L33,
  \dodoi{10.3847/2041-8213/ac9c63}

\bibitem[{{Lodato} {et~al.}(2009){Lodato}, {King}, \& {Pringle}}]{lodato09}
{Lodato}, G., {King}, A.~R., \& {Pringle}, J.~E. 2009, \mnras, 392, 332,
  \dodoi{10.1111/j.1365-2966.2008.14049.x}

\bibitem[{{Loeb} \& {Ulmer}(1997)}]{loeb97}
{Loeb}, A., \& {Ulmer}, A. 1997, \apj, 489, 573, \dodoi{10.1086/304814}

\bibitem[{{Lu} \& {Kumar}(2018)}]{lu18}
{Lu}, W., \& {Kumar}, P. 2018, \apj, 865, 128, \dodoi{10.3847/1538-4357/aad54a}

\bibitem[{{Magorrian} \& {Tremaine}(1999)}]{magorrian99}
{Magorrian}, J., \& {Tremaine}, S. 1999, \mnras, 309, 447,
  \dodoi{10.1046/j.1365-8711.1999.02853.x}

\bibitem[{{Mainetti} {et~al.}(2017){Mainetti}, {Lupi}, {Campana}, {Colpi},
  {Coughlin}, {Guillochon}, \& {Ramirez-Ruiz}}]{mainetti17}
{Mainetti}, D., {Lupi}, A., {Campana}, S., {et~al.} 2017, \aap, 600, A124,
  \dodoi{10.1051/0004-6361/201630092}

\bibitem[{{Marconi} {et~al.}(2004){Marconi}, {Risaliti}, {Gilli}, {Hunt},
  {Maiolino}, \& {Salvati}}]{marconi04}
{Marconi}, A., {Risaliti}, G., {Gilli}, R., {et~al.} 2004, \mnras, 351, 169,
  \dodoi{10.1111/j.1365-2966.2004.07765.x}

\bibitem[{{Merloni} \& {Heinz}(2008)}]{merloni08}
{Merloni}, A., \& {Heinz}, S. 2008, \mnras, 388, 1011,
  \dodoi{10.1111/j.1365-2966.2008.13472.x}

\bibitem[{{Merritt}(2013)}]{merritt13}
{Merritt}, D. 2013, Classical and Quantum Gravity, 30, 244005,
  \dodoi{10.1088/0264-9381/30/24/244005}

\bibitem[{{Metzger}(2022)}]{metzger22}
{Metzger}, B.~D. 2022, \apjl, 937, L12, \dodoi{10.3847/2041-8213/ac90ba}

\bibitem[{{Metzger} \& {Stone}(2016)}]{metzger16}
{Metzger}, B.~D., \& {Stone}, N.~C. 2016, \mnras, 461, 948,
  \dodoi{10.1093/mnras/stw1394}

\bibitem[{{Miles} {et~al.}(2020){Miles}, {Coughlin}, \& {Nixon}}]{miles20}
{Miles}, P.~R., {Coughlin}, E.~R., \& {Nixon}, C.~J. 2020, \apj, 899, 36,
  \dodoi{10.3847/1538-4357/ab9c9f}

\bibitem[{{Mockler} {et~al.}(2019){Mockler}, {Guillochon}, \&
  {Ramirez-Ruiz}}]{mockler19}
{Mockler}, B., {Guillochon}, J., \& {Ramirez-Ruiz}, E. 2019, \apj, 872, 151,
  \dodoi{10.3847/1538-4357/ab010f}

\bibitem[{{Mutlu-Pakdil} {et~al.}(2016){Mutlu-Pakdil}, {Seigar}, \&
  {Davis}}]{mutlu16}
{Mutlu-Pakdil}, B., {Seigar}, M.~S., \& {Davis}, B.~L. 2016, \apj, 830, 117,
  \dodoi{10.3847/0004-637X/830/2/117}

\bibitem[{{Nicholl} {et~al.}(2022){Nicholl}, {Lanning}, {Ramsden}, {Mockler},
  {Lawrence}, {Short}, \& {Ridley}}]{nicholl22}
{Nicholl}, M., {Lanning}, D., {Ramsden}, P., {et~al.} 2022, \mnras, 515, 5604,
  \dodoi{10.1093/mnras/stac2206}

\bibitem[{{Nixon} {et~al.}(2021){Nixon}, {Coughlin}, \& {Miles}}]{nixon21}
{Nixon}, C.~J., {Coughlin}, E.~R., \& {Miles}, P.~R. 2021, \apj, 922, 168,
  \dodoi{10.3847/1538-4357/ac1bb8}

\bibitem[{{Pasham} {et~al.}(2023){Pasham}, {Lucchini}, {Laskar}, {Gompertz},
  {Srivastav}, {Nicholl}, {Smartt}, {Miller-Jones}, {Alexander}, {Fender},
  {Smith}, {Fulton}, {Dewangan}, {Gendreau}, {Coughlin}, {Rhodes}, {Horesh},
  {van Velzen}, {Sfaradi}, {Guolo}, {Castro Segura}, {Aamer}, {Anderson},
  {Arcavi}, {Brennan}, {Chambers}, {Charalampopoulos}, {Chen}, {Clocchiatti},
  {de Boer}, {Dennefeld}, {Ferrara}, {Galbany}, {Gao}, {Gillanders}, {Goodwin},
  {Gromadzki}, {Huber}, {Jonker}, {Joshi}, {Kara}, {Killestein}, {Kosec},
  {Kocevski}, {Leloudas}, {Lin}, {Margutti}, {Mattila}, {Moore},
  {M{\"u}ller-Bravo}, {Ngeow}, {Oates}, {Onori}, {Pan}, {Perez-Torres}, {Rani},
  {Remillard}, {Ridley}, {Schulze}, {Sheng}, {Shingles}, {Smith}, {Steiner},
  {Wainscoat}, {Wevers}, \& {Yang}}]{pasham23}
{Pasham}, D.~R., {Lucchini}, M., {Laskar}, T., {et~al.} 2023, Nature Astronomy,
  7, 88, \dodoi{10.1038/s41550-022-01820-x}

\bibitem[{{Piran} {et~al.}(2015){Piran}, {Svirski}, {Krolik}, {Cheng}, \&
  {Shiokawa}}]{piran15}
{Piran}, T., {Svirski}, G., {Krolik}, J., {Cheng}, R.~M., \& {Shiokawa}, H.
  2015, \apj, 806, 164, \dodoi{10.1088/0004-637X/806/2/164}

\bibitem[{{Predehl} {et~al.}(2021){Predehl}, {Andritschke}, {Arefiev},
  {Babyshkin}, {Batanov}, {Becker}, {B{\"o}hringer}, {Bogomolov}, {Boller},
  {Borm}, {Bornemann}, {Br{\"a}uninger}, {Br{\"u}ggen}, {Brunner}, {Brusa},
  {Bulbul}, {Buntov}, {Burwitz}, {Burkert}, {Clerc}, {Churazov}, {Coutinho},
  {Dauser}, {Dennerl}, {Doroshenko}, {Eder}, {Emberger}, {Eraerds},
  {Finoguenov}, {Freyberg}, {Friedrich}, {Friedrich}, {F{\"u}rmetz},
  {Georgakakis}, {Gilfanov}, {Granato}, {Grossberger}, {Gueguen}, {Gureev},
  {Haberl}, {H{\"a}lker}, {Hartner}, {Hasinger}, {Huber}, {Ji}, {Kienlin},
  {Kink}, {Korotkov}, {Kreykenbohm}, {Lamer}, {Lomakin}, {Lapshov}, {Liu},
  {Maitra}, {Meidinger}, {Menz}, {Merloni}, {Mernik}, {Mican}, {Mohr},
  {M{\"u}ller}, {Nandra}, {Nazarov}, {Pacaud}, {Pavlinsky}, {Perinati},
  {Pfeffermann}, {Pietschner}, {Ramos-Ceja}, {Rau}, {Reiffers}, {Reiprich},
  {Robrade}, {Salvato}, {Sanders}, {Santangelo}, {Sasaki}, {Scheuerle},
  {Schmid}, {Schmitt}, {Schwope}, {Shirshakov}, {Steinmetz}, {Stewart},
  {Str{\"u}der}, {Sunyaev}, {Tenzer}, {Tiedemann}, {Tr{\"u}mper}, {Voron},
  {Weber}, {Wilms}, \& {Yaroshenko}}]{predehl21}
{Predehl}, P., {Andritschke}, R., {Arefiev}, V., {et~al.} 2021, \aap, 647, A1,
  \dodoi{10.1051/0004-6361/202039313}

\bibitem[{{Rees}(1988)}]{rees88}
{Rees}, M.~J. 1988, \nat, 333, 523, \dodoi{10.1038/333523a0}

\bibitem[{{Roth} {et~al.}(2016){Roth}, {Kasen}, {Guillochon}, \&
  {Ramirez-Ruiz}}]{roth16}
{Roth}, N., {Kasen}, D., {Guillochon}, J., \& {Ramirez-Ruiz}, E. 2016, \apj,
  827, 3, \dodoi{10.3847/0004-637X/827/1/3}

\bibitem[{{Ryu} {et~al.}(2020){Ryu}, {Krolik}, {Piran}, \& {Noble}}]{ryu20}
{Ryu}, T., {Krolik}, J., {Piran}, T., \& {Noble}, S.~C. 2020, \apj, 904, 99,
  \dodoi{10.3847/1538-4357/abb3cd}

\bibitem[{{Servin} \& {Kesden}(2017)}]{servin17}
{Servin}, J., \& {Kesden}, M. 2017, \prd, 95, 083001,
  \dodoi{10.1103/PhysRevD.95.083001}

\bibitem[{{Shankar} {et~al.}(2004){Shankar}, {Salucci}, {Granato}, {De Zotti},
  \& {Danese}}]{shankar04}
{Shankar}, F., {Salucci}, P., {Granato}, G.~L., {De Zotti}, G., \& {Danese}, L.
  2004, \mnras, 354, 1020, \dodoi{10.1111/j.1365-2966.2004.08261.x}

\bibitem[{{Shankar} {et~al.}(2009){Shankar}, {Weinberg}, \&
  {Miralda-Escud{\'e}}}]{shankar09}
{Shankar}, F., {Weinberg}, D.~H., \& {Miralda-Escud{\'e}}, J. 2009, \apj, 690,
  20, \dodoi{10.1088/0004-637X/690/1/20}

\bibitem[{{Shappee} {et~al.}(2014){Shappee}, {Prieto}, {Grupe}, {Kochanek},
  {Stanek}, {De Rosa}, {Mathur}, {Zu}, {Peterson}, {Pogge}, {Komossa}, {Im},
  {Jencson}, {Holoien}, {Basu}, {Beacom}, {Szczygie{\l}}, {Brimacombe},
  {Adams}, {Campillay}, {Choi}, {Contreras}, {Dietrich}, {Dubberley},
  {Elphick}, {Foale}, {Giustini}, {Gonzalez}, {Hawkins}, {Howell}, {Hsiao},
  {Koss}, {Leighly}, {Morrell}, {Mudd}, {Mullins}, {Nugent}, {Parrent},
  {Phillips}, {Pojmanski}, {Rosing}, {Ross}, {Sand}, {Terndrup}, {Valenti},
  {Walker}, \& {Yoon}}]{shappee14}
{Shappee}, B.~J., {Prieto}, J.~L., {Grupe}, D., {et~al.} 2014, \apj, 788, 48,
  \dodoi{10.1088/0004-637X/788/1/48}

\bibitem[{{Stone} {et~al.}(2019){Stone}, {Kesden}, {Cheng}, \& {van
  Velzen}}]{stone19}
{Stone}, N.~C., {Kesden}, M., {Cheng}, R.~M., \& {van Velzen}, S. 2019, General
  Relativity and Gravitation, 51, 30, \dodoi{10.1007/s10714-019-2510-9}

\bibitem[{{Stone} \& {Metzger}(2016)}]{stone16}
{Stone}, N.~C., \& {Metzger}, B.~D. 2016, \mnras, 455, 859,
  \dodoi{10.1093/mnras/stv2281}

\bibitem[{{Strubbe} \& {Quataert}(2009)}]{strubbe09}
{Strubbe}, L.~E., \& {Quataert}, E. 2009, \mnras, 400, 2070,
  \dodoi{10.1111/j.1365-2966.2009.15599.x}

\bibitem[{{Tonry} {et~al.}(2018){Tonry}, {Denneau}, {Heinze}, {Stalder},
  {Smith}, {Smartt}, {Stubbs}, {Weiland}, \& {Rest}}]{tonry18}
{Tonry}, J.~L., {Denneau}, L., {Heinze}, A.~N., {et~al.} 2018, \pasp, 130,
  064505, \dodoi{10.1088/1538-3873/aabadf}

\bibitem[{{van Velzen}(2018)}]{vanvelzen18}
{van Velzen}, S. 2018, \apj, 852, 72, \dodoi{10.3847/1538-4357/aa998e}

\bibitem[{{Volonteri} \& {Begelman}(2010)}]{volonteri10}
{Volonteri}, M., \& {Begelman}, M.~C. 2010, \mnras, 409, 1022,
  \dodoi{10.1111/j.1365-2966.2010.17359.x}

\bibitem[{{Wang} \& {Merritt}(2004)}]{wang04}
{Wang}, J., \& {Merritt}, D. 2004, \apj, 600, 149, \dodoi{10.1086/379767}

\bibitem[{{Will}(2012)}]{will12}
{Will}, C.~M. 2012, Classical and Quantum Gravity, 29, 217001,
  \dodoi{10.1088/0264-9381/29/21/217001}

\bibitem[{{Wu} {et~al.}(2018){Wu}, {Coughlin}, \& {Nixon}}]{wu18}
{Wu}, S., {Coughlin}, E.~R., \& {Nixon}, C. 2018, \mnras, 478, 3016,
  \dodoi{10.1093/mnras/sty971}

\bibitem[{{Yao} {et~al.}(2023){Yao}, {Ravi}, {Gezari}, {van Velzen}, {Lu},
  {Schulze}, {Somalwar}, {Kulkarni}, {Hammerstein}, {Nicholl}, {Graham},
  {Perley}, {Cenko}, {Stein}, {Ricarte}, {Chadayammuri}, {Quataert}, {Bellm},
  {Bloom}, {Dekany}, {Drake}, {Groom}, {Mahabal}, {Prince}, {Riddle},
  {Rusholme}, {Sharma}, {Sollerman}, \& {Yan}}]{yao23}
{Yao}, Y., {Ravi}, V., {Gezari}, S., {et~al.} 2023, arXiv e-prints,
  arXiv:2303.06523, \dodoi{10.48550/arXiv.2303.06523}

\bibitem[{{Zauderer} {et~al.}(2013){Zauderer}, {Berger}, {Margutti}, {Pooley},
  {Sari}, {Soderberg}, {Brunthaler}, \& {Bietenholz}}]{zauderer13}
{Zauderer}, B.~A., {Berger}, E., {Margutti}, R., {et~al.} 2013, \apj, 767, 152,
  \dodoi{10.1088/0004-637X/767/2/152}

\end{thebibliography}

\end{document}